\documentclass[10pt]{iopart}
\usepackage{hyperref}
\usepackage{iopams,natbib,graphicx}

\newcommand{\Mbh}{M_\bullet}
\newcommand{\trad}{P}
\newcommand{\tprec}{T_\mathrm{M}}
\newcommand{\trel}{T_\mathrm{rel}}
\newcommand{\rh}{r_\mathrm{m}}
\newcommand{\rcapt}{r_\bullet}
\newcommand{\Lcapt}{L_\bullet}
\newcommand{\Lcirc}{L_\mathrm{circ}}
\newcommand{\calD}{\mathcal D}
\newcommand{\calF}{\mathcal F}
\newcommand{\calR}{\mathcal R}
\newcommand{\calRcapt}{\mathcal R_\bullet}
\newcommand{\eaxi}{\epsilon}
\begin{document}

\title[Capture of stars by supermassive black holes]
{Rates of capture of stars by supermassive black holes in non-spherical galactic nuclei}

\author{Eugene Vasiliev}
\address{Lebedev Physical Institute, Moscow, Russia\\
E-mail: \url{eugvas@lpi.ru}\\
Received 28 February 2014, revised 13 May 2014\\
Accepted for publication 16 May 2014}
\begin{abstract}
We consider the problem of star consumption by supermassive black holes 
in non-spherical (axisymmetric, triaxial) galactic nuclei. 
We review the previous studies of the loss-cone problem and present 
a novel simulation method which allows to separate out the collisional 
(relaxation-related) and collisionless (related to non-conservation of 
angular momentum) processes and determine their relative importance 
for the capture rates in different geometries. 
We show that for black holes more massive than $10^7\,M_\odot$, 
the enhancement of capture rate in non-spherical galaxies is substantial, 
with even modest triaxiality being capable of keeping the capture rate 
at the level of few percent of black hole mass per Hubble time.
\end{abstract}

\section{Introduction}

The study of stellar dynamics around massive black holes begins in 1970s 
in application to (still hypothetical) intermediate-mass black holes in 
globular clusters \citep{BahcallWolf1976,FrankRees1976,LightmanShapiro1977,
CohnKulsrud1978}. 
The early studies established the importance of two-body relaxation which 
changes angular momenta of individual stars and drives them into the loss cone 
-- the region in phase space in which a star would be captured by the black 
hole in at most one orbital period. Later, the loss cone theory was applied 
to galactic nuclei \citep{DuncanShapiro1983,SyerUlmer1999,MagorrianTremaine1999,
WangMerritt2004}. 
If the black hole mass is not too large ($\lesssim 10^8\,M_\odot$), 
a star lost into the black hole could produce a tidal disruption flare 
\citep{Rees1988}, and several candidate events have been observed to date 
\citep[e.g.][]{StrubbeQuataert2009,Velzen2011}. 

To deliver a star into the loss cone, it is necessary to reduce its orbital 
angular momentum to a very small value. In a spherical geometry, this can only 
be achieved by two-body relaxation, but relaxation times in galactic nuclei are 
typically very long, especially in massive galaxies. 
The predicted rate of capture events is generally rather low, of order 
$10^{-5}\,M_\odot/$yr per galaxy, weakly depending on the black hole mass $\Mbh$. 
On the other hand, if a galaxy is not precisely spherically symmetric, then 
angular momenta of individual stars are not conserved, so that they may be 
driven into the loss cone by \textit{collisionless} torques (as opposed to 
\textit{collisional} two-body relaxation). This effect was recognized quite 
early \citep{NormanSilk1983,GerhardBinney1985}, and more recent studies have 
suggested that particular properties of triaxial galactic nuclei, namely the 
existence of a large population of centrophilic orbits, may keep the capture 
rate well above that of a similar spherical system \citep{MerrittPoon2004,
HolleySigurdsson2006, MerrittVasiliev2011}. These conclusions were obtained 
based on the properties of orbits, and not on the full-scale dynamical 
simulations, which are fairly non-trivial to conduct for such a problem.

There are several numerical methods that have been used for studying stellar 
systems with massive black holes. 
Fokker-Planck models are usually restricted to spherical \citep{CohnKulsrud1978} 
or at most axisymmetric \citep{Goodman1983,FiestasSpurzem2010,VasilievMerritt2013}
geometries, as are gaseous \citep{AmaroSeoaneFS2004} or Monte-Carlo 
\citep{ShapiroMarchant1978,FreitagBenz2002} models. 
$N$-body simulations \citep[e.g.][]{BaumgardtME2004,BrockampBK2011,FiestasPBS2012} 
do not have such limitation, but as we will show below, it is extremely hard to 
operate them in a regime with realistic proportion between collisional and 
collisionless effects. 
We have developed a new variant of Monte-Carlo code that is applicable in 
any geometry and have used it to study how the non-spherical effects change 
the rate of star captures by supermassive black holes.

We begin by reviewing the basic properties of orbits in galactic nuclei of 
various geometries in \S\ref{sec:orbits}. 
Then in \S\ref{sec:relaxation} we set up the loss-cone problem and consider 
the interplay between collisional and collisionless relaxation processes. 
\S\ref{sec:scaling} is devoted to order-of-magnitude estimates and scaling 
relations. In \S\ref{sec:montecarlo} we describe the novel Monte-Carlo method 
for simulating the dynamics of near-equilibrium stellar systems of arbitrary 
geometry, and in \S\ref{sec:results} apply this method for the problem 
of star capture by supermassive black holes in axisymmetric and triaxial 
galactic nuclei. \S\ref{sec:conclusions} presents the conclusions.

\section{Orbital structure of non-spherical galactic nuclei}  \label{sec:orbits}

Throughout this paper, we consider motion of stars in a time-independent potential, 
which is composed of a Newtonian potential of the supermassive black hole at origin 
(we ignore relativistic effects for the reasons described later) and the potential 
of extended distribution of stars:
\begin{equation}  \label{eq:potential}
\Phi(\boldsymbol{r}) \equiv -\frac{G \Mbh}{r} + \Phi_\star(\boldsymbol{r}) .
\end{equation}

It is clear that depending on the relative contribution of these two terms 
(or, rather, their derivatives that determine acceleration), there are two 
limiting regimes and a transition zone between them. Close to the black hole, 
the motion can be described as a perturbed Keplerian orbit. At larger radii, 
the black hole almost unnoticed by the majority of stars -- except for those 
with low angular momenta, which are able to approach close to the black hole;
nevertheless, even these orbits are mostly determined by the extended stellar 
distribution rather than the point mass. The boundary between the two cases 
is conveniently defined by the black hole influence radius $\rh$, which 
contains stars with the total mass $M_\star(<\rh) = 2\Mbh$. 
Another commonly used definition of the influence radius is $G\Mbh/\sigma^2$, 
where $\sigma$ is the stellar velocity dispersion; it is more readily computed 
from observations but has a more indirect meaning, as $\sigma$ is determined 
by the distribution of matter in the entire galaxy and not just close to the 
black hole, and may well be a function of radius itself. The two definitions 
of influence radius give the same value for a singular isothermal density 
profile, but may differ by a factor of few for less cuspy profiles 
\citep[Section 2.2]{MerrittBook}. We will use the first definition henceforth.
 
In general, the only classical integral of motion in a time-independent potential 
of arbitrary geometry is the total energy (per unit mass) of an orbit 
$E \equiv \Phi(\boldsymbol{r})+v^2/2$. Obviously, in the spherical and axisymmetric 
cases there exist additional integrals -- three and one components of the angular 
momentum vector $\boldsymbol{L}$, correspondingly. 
In the vicinity of the black hole, however, the orbits look like almost closed 
Keplerian ellipses, with the oscillations in radius between peri- and apoapses 
($r_-,r_+$) occuring on a much shorter timescale (radial period $\trad$) than 
the changes in the orientation of the ellipse (precession timescale). 
Under these conditions, one may use the method of orbit averaging 
\citep[e.g.][]{SridharTouma1999} to obtain another conserved quantity 
$I \equiv \sqrt{G \Mbh a}$, where $a$ is the semimajor axis%
\footnote{It is related to the radial action 
$I_r \equiv \frac{1}{\pi} \int_{r_-}^{r_+} v_r\,dr$ by $I=I_r+L$.}.
Thus the total Hamiltonian is split into the Keplerian and the 
perturbing part, each one being conserved independently.
The perturbing potential $\Phi_\star$, averaged over the radial 
oscillation period, becomes an additional integral of motion, the 
so-called secular Hamiltonian. 

The existence of this additional integral does not add new features in 
the spherical case, but is important in other cases. In particular, the motion 
in an axisymmetric nuclear star cluster becomes fully integrable, while in 
the weakly triaxial case another non-classical integral of motion can be shown 
to exist \citep{MerrittVasiliev2011}, so that the system is again fully integrable.
Empirical evidence shows that even for rather large deviation from spherical symmetry, 
orbits in the triaxial case remain regular in the vicinity of the black hole, 
if one neglects relativistic effects. 

\begin{figure}
$$\includegraphics[width=10cm]{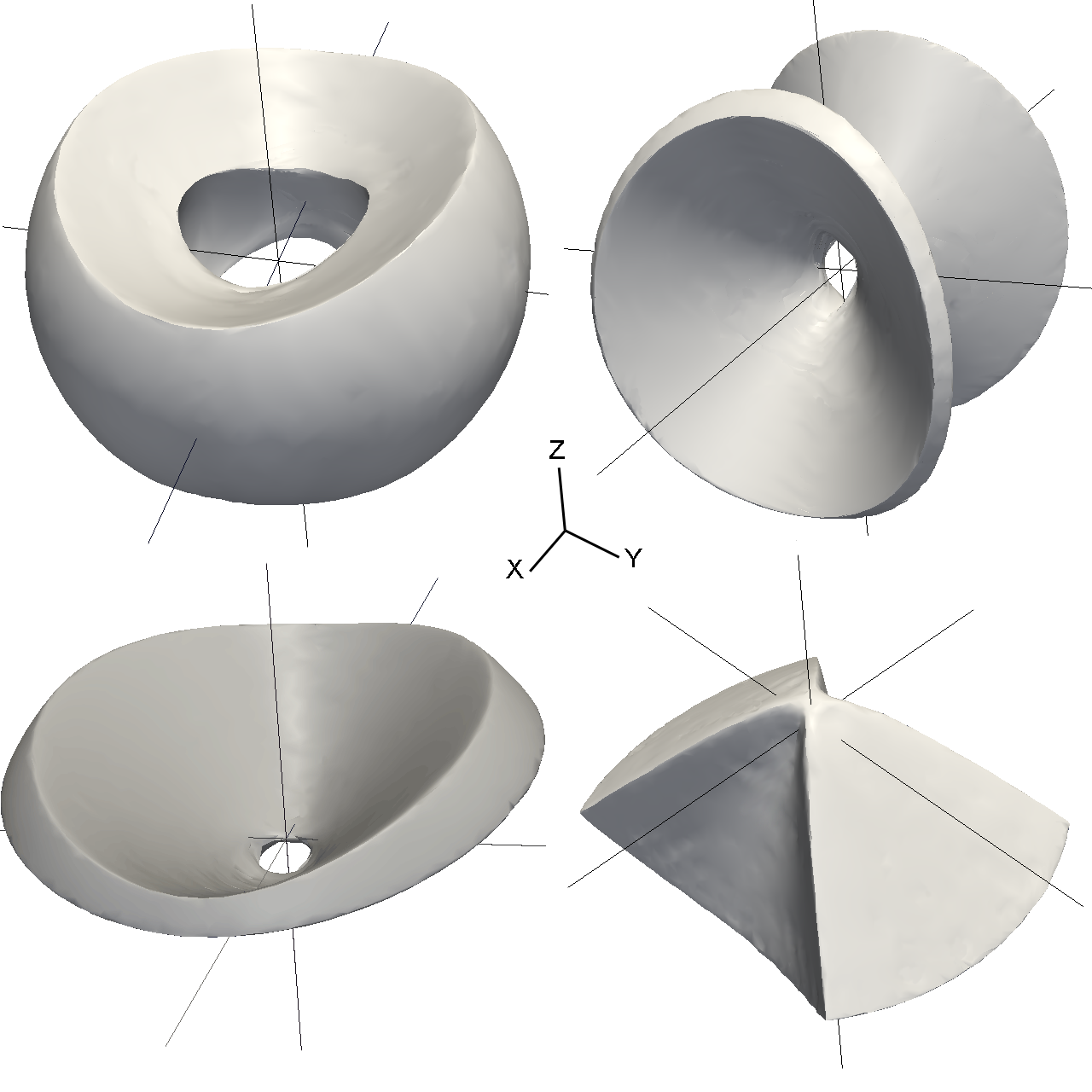}$$
\caption{Four major orbit families in a triaxial galactic nucleus:
short-axis tube (SAT, top left), (inner) long-axis tube (LAT, top right), 
saucer (bottom left), pyramid (bottom right). 
The long/intermediate/short axes are $x/y/z$. 
There are also outer long-axis tubes which look similar to SAT turned by 90 degrees 
(with $x$ axis going through the tube). 
In the axisymmetric case, only the two types of orbits on left panels exist. \protect\\
These orbits are found in the vicinity of the black hole (for $r \lesssim r_m$); 
outside the influence radius pyramid and saucer orbits (and some tubes as well) 
are largely replaced by chaotic orbits.
} \label{fig:orbits}
\end{figure}

In a generic triaxial potential there usually exist several types of tube orbits, 
which conserve the sign of one component of angular momentum, and box orbits, which 
do not have a definite sense of rotation around any axis \citep[][\S3.8]{BT2008}. 
The latter have stationary points, in which the velocity (and hence, the angular 
momentum) comes to zero, and are generally centrophilic, meaning that a star can 
pass at an arbitrarily small distance from the origin. Centrophilic orbits, by 
necessity, cannot conserve the sign of any component of angular momentum, however 
the opposite is not true -- there exist various non-tube resonant orbit families 
which avoid coming too close to the center. These orbits play an important role 
in the structure of triaxial galaxies with density cusps and/or central black holes 
\citep{MerrittValluri1999}, as the regular box orbits are destroyed by the sharp 
gradients in the potential near origin and replaced by chaotic orbits and resonant 
orbit families. Thus the centrophilic orbits outside a few influence radii and 
up to a radius containing $M_\star \sim 10^2\,\Mbh$ are generally chaotic 
\citep{ValluriMerritt1998}, although in some special cases the motion remains fully 
integrable even in the presence of the black hole \citep{SridharTouma1997}. 
On the other hand, the motion inside $\rh$ is generally regular, with the box orbits 
being replaced by a related family of pyramid orbits
\citep[][Fig.11a]{MerrittValluri1999}. These are also centrophilic, and as such, 
are natural candidates for the capture by the black hole \citep{MerrittVasiliev2011}. 
Tube orbits also exist near the black hole, and a special class of short-axis tubes 
has a noticeably conical shape and are called saucers \citep{SambhusSridhar2000}. 
Figure~\ref{fig:orbits} shows the four major classes of orbits in a triaxial nucleus, 
of which two exist also in the oblate axisymmetric case. 
In an even more general case without a triaxial symmetry (for instance, if the black 
hole is displaced from the center of stellar potential), there are other classes of 
orbits \citep[e.g.\ lens orbits, see][]{SridharTouma1999}, some of them also being 
centrophilic. In what follows, we consider systems with at least triaxial symmetry.

Given the non-conservation of angular momentum in non-spherical potentials, it is 
important to know the amplitude and timescale of its variation. 
Let us describe the deviation from spherical symmetry by a parameter $\eaxi$, 
roughly equal to $1-x/z$, where $x$ and $z$ are the major and minor axes
(more precise definition is given by Eq.11b in \citet{VasilievMerritt2013}, where 
it is called $\calR_\mathrm{sep}$).
For a triaxial system, there is another parameter $T=(x^2-y^2)/(x^2-z^2)$
which describes the degree of triaxiality ($y$ is the intermediate axis). 
It ranges from $T=0$ for oblate to $T=1$ for prolate axisymmetric models; in the rest 
of the paper we will assume $T=0.5$ (maximal triaxiality), having in mind that 
the importance of triaxial effects decreases as the system approaches axisymmetry.
Let $L_\mathrm{circ}(E)$ be the angular momentum of a circular orbit with the given 
energy, and define the dimensionless squared angular momentum 
$\calR \equiv L^2/\Lcirc^2$.
It turns out that for tube orbits with high enough $\calR$ the variation in $\calR$ 
is rather small, but for saucer and pyramid orbits, both of which appear at 
$\calR \lesssim \eaxi$, this variation can be by order of $\calR$ itself. 
(Note, however, that in the axisymmetric case the magnitude of angular momentum is 
bounded from below by its conserved component $L_z$). 
The timescale for such variation is of order $\eaxi^{-1/2}$ times the precession time
$\tprec = P\,\Mbh/M_\star(a)$ \citep{VasilievMerritt2013}.
The above estimates were valid for the motion in the sphere of influence; 
outside $\rh$ the chaotic centrophilic orbits also typically have 
$\calR \lesssim \eaxi$ and the variation of angular momentum by order of itself 
occurs on the timescale of radial period $\trad$. 

There is ample evidence that a substantial fraction of elliptical galaxies and bulges 
of disk galaxies might be triaxial \citep{TremblayMerritt1996, MendezAbreu2010}, with 
the minor-to-major axis ratio in the range $0.7-0.85$ and a large spread in the 
triaxiality parameter. Nuclear star clusters also are known to exhibit significant 
departures from spherical symmetry \citep[see][for the case of Milky Way]{Schoedel2014}, 
which is also a natural consequence of one of the proposed scenarios of their formation 
\citep{Antonini2014}. 
Even if the mass distribution in the nuclear star cluster itself is spherical, 
the distortion of the gravitational potential by asymmetries on larger scales (such as 
nuclear bars) would still create the centrophilic orbit families.
We will take the axis ratio $z/x=0.8$ and triaxiality $T=0.5$ as our fiducial parameters, 
but we note that the capture rates do not depend substantially on these values.

\section{Collisional and collisionless relaxation and the loss-cone problem}  \label{sec:relaxation}

The black hole tidally disrupts or directly captures stars which pass at a distance 
smaller than a critical radius $\rcapt$, or, equivalently, pass the periapse while 
having orbital angular momentum smaller than $\Lcapt$. The former fate occurs for 
main-sequence stars if $\Mbh \lesssim 10^7-10^8\,M_\odot$, depending on the mass of 
the star \citep[see Fig.5 in][]{FreitagBenz2002}, and for giant stars even for much 
larger $\Mbh$; the latter case applies to compact stars, stellar-mass black holes, 
or for the largest $\Mbh$. We shall consider only non-rotating black holes; 
see the paper by L.Dai in this volume for the problem of tidal disruption of stars 
by Kerr black holes. The critical angular momentum is
\begin{equation}  \label{eq:Lcapt}
\Lcapt = \sqrt{2G\Mbh \rcapt};\quad \rcapt \equiv 
\mathrm{max}\left[ \frac{8G\Mbh}{c^2}, 
\left(\eta^2\frac{\Mbh}{m_\star}\right)^{1/3} R_\star\right] ,
\end{equation}
where $m_\star$ and $R_\star$ are the mass and radius of the star and 
the coefficient $\eta\sim 1$ depends on the internal structure of the star. 
The region of phase space with $L<\Lcapt$ is called the loss cone.

A detailed review of the loss cone theory is given by \citet{Merritt2013}; 
here we remind several key points.
Substantial changes in both energy and angular momentum distributions occur 
on the timescale of relaxation time $\trel(E)$, but the relaxation in angular 
momentum is more important for the overall rate of capture \citep{FrankRees1976}, 
since near the loss cone boundary a comparatively small change in $L$ can drive 
a star into the black hole. 
The relaxation in $E$ would produce a $\rho\propto r^{-7/4}$ cusp 
\citep{BahcallWolf1976} in a time $\sim0.1\trel(\rh)$; since many galaxies are not 
that dynamically old, we will not assume this particular density profile 
and let it to be an unspecified function. The isotropic distribution function of 
stars in energy $\overline f(E)$ is given by the Eddington formula, while the 
density of states $g(E)=4\pi^2\,P(E)\,L_\mathrm{circ}^2(E)$ provides the link 
between $f$ and the number of stars in energy interval $dE$: 
$\overline N(E)\, dE = \overline f(E)\, g(E)\, dE$ \citep[][Eq.4.46, 4.56]{BT2008}.
In what follows, we will use interchangeably the energy $E$ or the semimajor axis $a$
(for a non-Keplerian potential it should be more appropriately called the radius of 
a circular orbit with the given energy). 

In the spherical case, the only mechanism producing changes in angular momentum 
is the two-body relaxation. It can be substantially enhanced if there are massive 
perturbing objects such as molecular clouds \citep{PeretsHA2007}, but the basic 
physical effect remains the same. Resonant relaxation also enhances the exchange 
of angular momentum between stars \citep{RauchTremaine1996, HopmanAlexander2006}, 
but has a rather moderate effect on the overall rate of capture, so we also ignore 
it in the discussion. Relativistic effects play an important role close to the 
black hole \citep[C.Will, this volume]{MerrittAMW2011}, but since the bulk of 
captured stars come from radii comparable to $\rh$, relativity can be neglected for 
them. On the other hand, in non-spherical systems the angular momentum changes 
due to both two-body relaxation and torques from non-spherical mass distribution, 
the interplay between these two effects being quite complex.

In the case of spherical symmetry, the main parameter that determines the 
flux of stars into the loss cone is the ratio $q$ of mean-square change in angular 
momentum per one radial period to the width of the loss cone:
\begin{equation}  \label{eq:q_spher}
q(E) \equiv \frac{\trad(E) \calD(E)}{\calRcapt(E)} \;,\quad
\calRcapt \equiv \frac{\Lcapt^2}{\Lcirc^2(E)} = \frac{2\rcapt}{a(E)}, 
\end{equation}
where $\calD(E)$ is the limiting value of diffusion coefficient in $\calR$ for 
$\calR\to0$, roughly equivalent to $\trel^{-1}$. 
This quantity determines the boundary condition for the diffusion in $\calR$ 
in the following way \citep{LightmanShapiro1977}. When $q \ll 1$, the relaxation 
is rather weak and as soon as the star reaches the boundary of the loss cone, 
it is captured within one orbital period; thus the density of stars inside the 
loss cone is zero -- this is called the empty loss cone regime. 
On the other hand, if $q \gg 1$, the changes in angular momentum occur so rapidly 
that in the time-averaged sense, the loss cone remains fully populated: 
the stars with $\calR<\calRcapt$ are still being captured once per radial period, 
but this occurs only when the star actually passes the periapse, and for the rest 
of its orbit it can stay inside the loss cone without impunity. 
If we introduce the time-averaged value of distribution function in the loss cone 
as $f_\mathrm{LC}\equiv f(E,\calRcapt)$, then the rate of capture is given by 
the number of stars inside the loss cone divided by radial period:
\begin{equation}  \label{eq:flux_full_lc}
\calF_\mathrm{full\ LC}(E)\, dE \equiv 
  \frac{\calRcapt(E)\, f_\mathrm{LC}(E)\,g(E)\, dE}{\trad(E)} = 
  8\pi^2\,G\Mbh\rcapt\,f_\mathrm{LC}(E) \,dE.
\end{equation}

In the spherical case, the boundary condition fully determines the steady-state 
solution of the diffusion equation: assuming a particular form of isotropized 
(averaged over angular momenta) distribution function of stars $\overline f(E)$, 
the actual two-dimensional function $f(E, \calR)$ has a nearly logarithmic 
profile in $\calR$, and the rate of capture is \citep{VasilievMerritt2013}
\begin{equation}  \label{eq:flux_spherical}
\calF_\mathrm{spher}(E)\, dE \approx 
\frac{\calD(E)\, \overline f(E)\, g(E)\, dE}{\alpha + \ln(1/\calRcapt) - 1} , 
\quad \alpha \approx (q^4+q^2)^{1/4} .
\end{equation}

It is easy to see that in the full loss cone limit ($q\gg1$), this quantity 
tends to (\ref{eq:flux_full_lc}), and $f_\mathrm{LC}\approx \overline f(E)$.
Close to the black hole the loss cone is empty, and further out it may -- or may 
not -- reach a full loss cone regime; for $\Mbh \gtrsim 10^8\,M_\odot$ in 
lower-density galactic nuclei this never happens, while for smaller $\Mbh$ 
the transition between empty and full loss cone regimes may lie around $\rh$.

For non-spherical system, the treatment of two-body relaxation is more difficult 
and has only been carried out rigorously in the axisymmetric case inside the sphere 
of influence \citep{VasilievMerritt2013}, where the dynamics is fully integrable.
The main conclusion is that the steady-state solution at $\calR\gtrsim \eaxi$ 
is still well described by a logarithmic profile, but at smaller $\calR$ there 
exists a more extended ``loss wedge'' \citep{MagorrianTremaine1999}, in which 
the orbits on average have $\calR>\calRcapt$ but may get into the loss cone 
(below $\calRcapt$) due to torques from non-spherical potential, roughly after 
one precession time $\tprec$. 
The volume of the loss wedge is larger than the loss cone by a factor 
$\sim \sqrt{\eaxi/\calRcapt}$, and the boundary condition corresponds to a full 
loss cone regime (i.e.\ the distribution function inside the loss cone is almost 
the same as on the boundary of the loss wedge). In other words, the quantity 
$q_\mathrm{axi}$, analogous to (\ref{eq:q_spher}), is $\gg 1$ for most orbits, 
because the change of angular momentum due to precession during one radial period 
is typically larger than the size of the loss cone.
However, now the global shape of the solution is not simply determined by 
the boundary condition, but rather by the average two-body relaxation rate
at the given energy: the non-spherical torques cannot bring an orbit all the way 
from $\calR\sim 1$ to $\calRcapt$ --  this still remains the duty of relaxation. 
Overall, the larger size of the loss wedge increases the steady-state 
capture rate by a factor of few, compared to the spherical case, but only 
in the regime when the spherical system would have empty loss cone. 
Roughly speaking, the star only needs to get down to $\calR \sim \eaxi$ 
by collisional diffusion, and then it will be driven into the loss cone 
by collisionless effects; but as the dependence of the flux on the 
effective size of the loss region is logarithmic (\ref{eq:flux_spherical}), 
this does not greatly increase the loss rate.

A similar picture should be describing triaxial systems; however, for them, 
and to a lesser extent for axisymmetric systems, another effect may be more 
important, namely the draining of the loss region. 
This term refers to the gradual depletion of the region in phase space 
populated by orbits that may attain angular momentum lower than $\Lcapt$ 
at some point of their evolution in the smooth stationary potential 
(i.e.\ in the absense of relaxation).
In the triaxial case, this region contains the entire population of 
centrophilic orbits (which constitute approximately a fraction $\eaxi/4$ of 
total mass), whether these are regular pyramids inside $\rh$ or chaotic orbits 
further out. 
The lifetime of orbits in the loss region is much longer in the triaxial case; 
it is roughly given by \citep[][Eq.117]{MerrittVasiliev2011}:
\begin{equation}  \label{eq:tdrain_tri}
T_\mathrm{drain,triax} \sim \frac{\eaxi}{4}\frac{\trad(E)}{\calRcapt(E)} =
 \frac{\eaxi\,g(E)}{32\pi^2\,G\Mbh\rcapt} \approx
10^{10}\, \eaxi \left(\frac{a}{10\;\mathrm{pc}}\right)^{5/2} 
\left(\frac{\Mbh}{10^8\;M_\odot}\right)^{-3/2} \;\mathrm{yr},
\end{equation}
the latter equation used the radius of direct capture and not tidal disruption 
and is valid for $a \lesssim \rh$. 
An analogous expression for draining of chaotic orbits in the axisymmetric case 
\citep[][Eq.65]{Merritt2013} gives 
\begin{equation}  \label{eq:tdrain_axi}
T_\mathrm{drain,axi} \sim 10^8 \eaxi^{1/2} 
\left(\frac{a}{10\;\mathrm{pc}}\right)^{2}
\left(\frac{\Mbh}{10^8\;M_\odot}\right)^{-1} \;\mathrm{yr}.
\end{equation}

Thus it is clear that the timescale for draining the loss region can be quite 
long, especially in the triaxial case; during this time, the capture rate at 
a given energy is roughly given by the expression (\ref{eq:flux_full_lc}) for 
the full loss cone.
The necessary conditions for this are that  (i) the initial distribution of stars 
in $\calR$ is close to uniform (isotropic), and (ii) the typical change in angular 
momentum during one radial period is larger than the loss cone size.
The first condition is the most neutral, if not natural, choice for the 
velocity distribution function, although it may be substantially overestimating 
the capture rate if the low angular momentum part of the phase space is initially 
depleted due to the slingshot ejection of stars by a binary supermassive black 
hole \citep{MilosMerritt2003}.
The second condition is satisfied for most orbits which are not too close to 
the black hole (essentially most orbits that contribute to the total flux), 
as shown by \citet[][Eq.55]{MerrittVasiliev2011} and 
\citet[][Eq.59]{VasilievMerritt2013}.

It is also worth noting that the above discussion assumed a steady-state solution 
for the distribution function (to the extent that we ignore the loss of stars 
into the black hole, which is reasonable on the Hubble timescale). 
A time-dependent solution to the Fokker-Planck equation may give the loss rate 
higher than the steady-state by a factor of few \citep{MilosMerritt2003,
VasilievMerritt2013}, or, conversely, lower \citep{MerrittWang2005}, depending 
on the initial conditions. 
There are two separate factors that complicate matters in the non-stationary case.
The first is that if the relaxation time is much longer than the Hubble time, 
then the gradient of the distribution function near the capture boundary, which 
determines the capture rate in the empty loss cone regime, is not uniquely 
determined by the isotropized distribution function $\overline f(E)$, as 
in the steady state (Eq.~\ref{eq:flux_spherical}). Rather, it depends on the prior 
evolution and, ultimately, the initial state.
Figure~13 in \citet{VasilievMerritt2013} shows that under the assumption of 
initially isotropic distribution of stars, the time-dependent capture rate for 
a spherical galaxy may be $2-3$ times higher than the steady-state rate for the 
most massive black holes. On the other hand, these two rates are roughly equal 
for $\Mbh\lesssim 10^8\,M_\odot$, when the relaxation is fast enough to establish 
a nearly steady-state profile. 
The second factor, more important for non-spherical systems, is related to 
the collisionless draining of the loss region, which could increase the captured 
mass by more than an order of magnitude compared to the steady-state solution.
However, one should keep in mind that both factors sensitively depend on 
the initial population of stars in and near the loss region 
(i.e., on the details of galaxy formation), while the steady-state rates, 
which differ only by a factor of two between spherical and axisymmetric 
systems, provide a more robust estimate.

\section{Scaling laws}  \label{sec:scaling}

We can obtain a very rough estimate of the total steady-state capture rate 
for a spherical galaxy by noting that $\calF(E)$ peaks around $E\sim \Phi(\rh)$, 
so that if the system is in the empty loss cone regime at this energy,
then the total flux is
\begin{equation}  \label{eq:Ftotal_spher}
\calF_\mathrm{total} \sim \frac{\Mbh}{0.1\trel\,\ln\calRcapt^{-1}} 
\sim 10^{-6}\, \left(\frac{\Mbh}{10^8\,M_\odot}\right)^2
  \left(\frac{\sigma}{200\,\mbox{km/s}}\right)^{-3}
  \left(\frac{\rh}{100\,\mbox{pc}}\right)^{-3}  \; M_\odot/\mathrm{yr}.
\end{equation}

To arrive at a single-parameter formula, we use the $\Mbh-\sigma$ relation, 
which is typically written as
\begin{equation}  \label{eq:M_sigma}
\log(\Mbh/M_\odot) = \alpha + \beta \log(\sigma/200\,\mbox{km\,s}^{-1})\;,
\end{equation}
with $\alpha\approx 8$, $\beta\approx 4.5$ being the typical values found in 
the literature \citep[e.g.][]{FerrareseMerritt2000, Gebhardt2000, Gultekin2009}.
The influence radius then scales as 
\begin{equation}  \label{eq:rinfl_M}
\rh \approx r_0 (\Mbh/10^8\,M_\odot)^{0.56}\mbox{ pc},
\end{equation}
with the parameter $r_0$ depending on the density profile. 
Taking $\sigma$ to be the line-of-sight velocity, averaged over the effective radius, 
we obtain $r_0=45$~pc for a \citet{Dehnen1993} model with a $\gamma=1$ cusp. 
Then the estimate (\ref{eq:Ftotal_spher}) gives
$\calF_\mathrm{total} \sim 10^{-5} (\Mbh/10^8\,M_\odot)^{-0.3} M_\odot/$yr 
for more massive galaxies which are in the empty loss cone regime, 
comparable with more accurate calculations \citep[][Fig.13]{VasilievMerritt2013}.

For an axisymmetric galaxy, the steady-state capture rate is only moderately 
larger than for a spherical one, and have roughly the same functional dependence 
on galaxy parameters \citep{MagorrianTremaine1999}. The statement that axisymmetric 
galaxies have much higher capture rates is often attributed to this paper, but 
in fact a more careful reading reveals that the effect of axisymmetry increases 
the capture rates due to relaxation only by a factor of two or so. 
This ignores the draining of the loss wedge, which is justified since the draining 
time is substantially shorter than the Hubble time at the radii of interest,  
as is the case for most realistic situations: substituting (\ref{eq:rinfl_M}) into 
(\ref{eq:tdrain_axi}), we obtain that $T_\mathrm{drain,axi} \lesssim 10^9$~yr at 
the radius of influence.
On the other hand, the mass of stars in the loss wedge, and hence their draining 
rate, is roughly proportional to the mass of the black hole, while the capture rate 
due to relaxation scales more slowly with $\Mbh$. Hence, \citet{MagorrianTremaine1999} 
suggest that for the most massive galaxies it is indeed the draining of the loss 
wedge that is the main source of captured stars, and the capture rate exceed the 
steady-state value by orders of magnitude, which is apparent in comparing their 
figures 3 and 5.
Thus their study presented two very different conclusions about capture rates in 
axisymmetric galaxies, without a clear explanation of the discrepancy. 
It is worth reiterating that the first one, about a moderate increase in the 
relaxation flux, is more robust, as it does not depend on the initial conditions, 
while the second one, about a much higher capture rate due to draining, is based on 
the assumption that the loss wedge was fully populated at the beginning of evolution.
Both conclusions were confirmed by \citet{VasilievMerritt2013} using a more 
elaborate calculation.

For a triaxial galaxy, however, the draining times for most galaxies are comparable 
to or exceed the Hubble time, and the capture rate at a given energy is comparable 
to the full loss cone rate (\ref{eq:flux_full_lc}) if 
$T_\mathrm{drain,triax}(E) > T_\mathrm{Hubble}$ (again under the assumption of 
initially fully populated loss region). 
We may make a crude estimate by integrating the expression for the full loss 
cone flux from the critical energy up to infinity, where 
the critical energy $E_\mathrm{drain}$ is given by the condition that the
draining time (\ref{eq:tdrain_tri}) is equal to the Hubble time. 
Hence, the total capture rate is
\begin{equation}  \label{eq:Fdrain}
\calF_\mathrm{total} = \int_{E_\mathrm{drain}}^0 
  \frac{dE\,f(E)\,g(E)\,2G\Mbh\rcapt}{P(E)\,\Lcirc^2(E)} = 
  8\pi^2 G\Mbh\rcapt \int_{E_\mathrm{drain}}^0 f(E)\,dE .
\end{equation}

From (\ref{eq:tdrain_tri},\ref{eq:rinfl_M}) we see that, evaluated at the radius 
of influence, $T_\mathrm{drain}\sim 4\times10^{11}\,\eaxi$~yr, so that unless 
the triaxiality is very small, most of the draining flux comes from inside $\rh$.
We may make the above estimate more quantitative by taking the expressions for 
$f(E)$ and $g(E)$ in the Newtonian potential 
\citep[][Eqs.3.49, 3.51, 5.182]{MerrittBook}. This yields 
\begin{equation}  \label{eq:Ftotal_triax}
\calF_\mathrm{total} \sim 10^{-4}\, 
  \left(\frac{\Mbh}{10^8\,M_\odot}\right)^{\frac{14-3\gamma}{5}}
  \left(\frac{\rh}{100\,\mbox{pc}}\right)^{-(3-\gamma)}
  \left(\frac{t/\eaxi}{10^{10}\,\mbox{yr}}\right)^{-\frac{2\gamma-1}{5}} 
  \; M_\odot/\mathrm{yr} .
\end{equation}

The capture rate increases approximately linearly with $\Mbh$, a situation rather 
different from the spherical case. For a $\gamma=1$ cusp, the present-day flux is 
$\sim 4\times 10^{-4}\, \eaxi^{1/5} (\Mbh/10^8\,M_\odot)^{1.1}\;M_\odot \mathrm{yr}^{-1}$.
Note that, as the more bound orbits are drained more quickly, the total rate 
declines with time, although rather slowly. Even with such a high flux, it is 
not quite correct to say that the capture rate is comparable or exceeds the full 
loss cone rate, as the latter quantity is poorly defined by itself: the undepleted 
full loss cone draining rate would \textit{increase} with increasing $|E|$, 
producing a divergent overall total flux \citep[][Chapter 6]{MerrittBook}. 
A rough estimate obtained for a singular isothermal sphere ($r^{-2}$ density cusp), 
taking the full loss cone rate at the radius of influence, produces another two 
orders of magnitude larger flux \citep{ZhaoHR2002}.

The scaling relation (\ref{eq:Ftotal_triax}) is somewhat different from the one 
obtained by \citet{MerrittPoon2004} based on their model of scale-free 
$\gamma=1$ cusp with $x:y:z=1:0.8:0.5$. 
There are several reasons for this. First, their models, having a power-law 
density profile, could not extend to infinity, so were cut off at a few influence 
radii; second, their plots of capture rate versus energy used the definition of 
potential without the contribution of black hole, which is strictly valid only 
well outside influence radius. In short, there is no range of radii such that 
$\rh \ll r \ll r_\mathrm{cutoff}$, so it is hard to interpret their power-law fits.
In particular, the dashed line on their Fig.3, marking the full loss cone flux, 
is correct for $r\gg \rh$, and it is indeed apparent that for large energies the 
points start to follow that asymptote, instead of the power-law fit for smaller 
energies. On the other hand, a correct expression for the full loss cone flux 
valid both inside and outside has a peak at roughly $E=\Phi(\rh)$ and declines 
towards lower energies as $(-E)^{-1/2}$; however, as it extends to arbitrarily 
negative values of $E$, the overall flux would diverge at low, not at high $E$ 
as they claim. This divergence is removed by imposing a lower limit at energy 
$E_\mathrm{drain}$, which introduces the time dependence in the total flux 
that was derived in the above formulae. Lastly, their normalization is somewhat 
lower because the capture radius was taken to be at one Schwarzschild radius 
instead of four. Despite these complications, the basic statement, that the 
draining rate may be equal to the full loss cone rate during a Hubble time, 
remains valid if we take into account the proper cutoff at energies below 
$E_\mathrm{drain}$, and even the numerical estimates are correct to within 
an order of magnitude. 

We now discuss scaling relations that may facilitate numerical study of loss rates.
State-of-the-art $N$-body codes on the present-day hardware can simulate systems 
with $\sim10^5$ stars for $\sim10^5$ dynamical times, which takes months to years 
of wall-clock time (D.Heggie, priv.comm.). A simulation of even the relatively 
low-mass Milky Way nuclear star cluster would need to follow some $10^7$ stars 
for $10^5$ dynamical times, clearly beyond the reach of direct $N$-body simulations 
in the near future. 
However, we may design (almost) equivalent systems with suitably scaled parameters.

We start from spherical case, in which the important quantity that needs to be 
preserved by the scaling is the radius (or energy) of transition between 
empty and full loss cone regimes, i.e.\ at which $q=1$ (Eq.\ref{eq:q_spher}). 
Then it is clear that we need to change $\calRcapt$ and $\calD$ by the same factor
$\beta$. The flux in the full loss cone regime (Eq.\ref{eq:flux_full_lc}) will be 
also multiplied by $\beta$, and the flux in the empty loss cone regime 
(Eq.\ref{eq:flux_spherical} with $\alpha\to 0$) will be multiplied by almost 
the same factor, except for the change in the logarithmic factor in the denominator.
The total simulation time, in units of dynamical time, should then be scaled by 
$\beta^{-1}$. 

In the axisymmetric case the scaling is largely the same, if we ignore the draining 
of the loss wedge; furthermore, in this case $\calRcapt$ under the logarithm 
in the denominator is replaced with an effective constant which depends only on 
$\eaxi$ \citep{VasilievMerritt2013}. The draining, however, scales differently 
with $\beta$, so we may only ignore its contribution if the scaled draining time 
is less than the scaled simulation time.

In the triaxial case, the flux from draining of centrophilic orbits cannot be 
ignored as it gives the dominant contribution to the total capture rate. We need to 
preserve the ratio of draining time (\ref{eq:tdrain_tri}) to the simulation time, 
which, fortunately, means that both should change by the same factor $\beta^{-1}$. 
Thus the relative importance of collisional and collisionless capture rate will 
not change under the same scaling of $\calD\propto \beta, \calRcapt\propto \beta, 
t_\mathrm{sim}\propto \beta^{-1}$. 

Unfortunately, even with this scaling it is not possible to design a feasible 
$N$-body simulation for most of galaxies: since the relaxation rate 
$\calD \propto (\ln N)/N$, and we need to consider galaxies with $N\gtrsim 10^{10}$, 
$\beta$ should be of order $10^4$. Thus instead of a system which is $\sim 10^4$ 
dynamical times old at $\rh$, we get a system which has an age of $\sim 1$ 
dynamical time, and a capture radius which is a substantial fraction of $\rh$; 
clearly, such an extrapolation is too brave. 
In addition, for the system with a binary black hole the scaling is impossible to 
follow, as the stars are lost due to three-body scattering and not to the capture, 
so that the relevant radius $\rcapt$ is proportional to the size of the binary 
orbit and cannot be changed at will. In such case, increasing the relaxation rate 
without adjusting the loss cone size moves the system more into a full loss cone 
regime, and decreases the relative importance of draining.

\section{Monte-Carlo treatment of relaxation in arbitrary geometry}  \label{sec:montecarlo}

The complexity of interplay between collisional relaxation and collisionless changes 
in angular momentum makes a direct solution of Fokker-Planck equation impractical in 
the general case of triaxial system with a mixture of regular and chaotic orbits. 
On the other hand, direct $N$-body simulations are presently incapable of achieving 
the necessary regime of relatively weak two-body relaxation, compared to the 
collisionless effects. 

To address this problem, we have developed a novel Monte-Carlo method. 
The idea is to follow the motion of sample orbits in a smooth, non-spherical potential 
while adding a random perturbation to the equations of motion, which would mimick 
two-body relaxation with an \textit{adjustable} magnitude (unrelated to the actual 
number of test particles). 
The approach is actually not so novel, dating back to \citet{SpitzerHart1971}, who 
used it for a spherical system with diffusion coefficients being computed under an 
assumption of Maxwellian distribution of background (field) star velocities. 
The perturbations were applied directly to the velocity of the test star, its motion 
being numerically integrated in the given potential. 
Later on, the idea was modified in \citet{SpitzerShapiro1972} by using orbit-averaged 
diffusion coefficients in energy, and further developed by \citet{ShapiroMarchant1978} 
and \citet{DuncanShapiro1983} to study star clusters around supermassive black holes, 
using two-dimensional orbit-averaged diffusion in $\{E,L\}$ space with a special 
treatment of loss-cone ensuring a correct form of the boundary condition. 
The diffusion coefficients were recomputed regularly, using the distribution function 
of field stars equal to the isotropized distribution of test stars.
At the same time, another branch of Monte-Carlo methods was developed by 
\citet{Henon1971}, in which the perturbations were computed between pairs of stars, 
thus equating the test and the field star population and ensuring conservation of 
global integrals of motion in each interaction. This method is employed in several 
presently used Monte-Carlo codes for the evolution of spherically symmetric star 
clusters \citep[e.g.][]{Giersz1998, JoshiRPZ2000, FreitagBenz2002}. Unfortunately, 
it is difficult to generalize it for non-spherical systems, while the Spitzer's 
formulation is relatively straightforward to apply in any geometry.

The detailed description of the new Monte-Carlo code is given elsewhere 
\citep{Vasiliev2014}; here we outline its main features.
As in all previous studies, we consider scattering of test particles by an isotropic 
spherically symmetric distribution of field particles, which are described 
by a distribution function $\overline f(E)$ consistent with the spherical part of 
the potential in which the stars move. 
The scattering is described in terms of local (position-dependent) velocity diffusion 
coefficients $\langle \Delta v_\| \rangle, \langle \Delta v^2_\| \rangle, 
\langle \Delta v^2_\bot \rangle$ \citep[e.g.][Eqs.5.23, 5.55]{MerrittBook},
which represent mean and mean-squared changes in velocity per unit time.
The trajectory of a test star is integrated numerically in the given smooth potential, 
and the perturbations to velocity are applied after each timestep of the ODE integrator, 
typically several tens of steps per radial period, using locally computed drag and 
diffusion coefficients. 
We have checked that the prescription for adding velocity perturbation gives correct 
rates of orbit-averaged diffusion coefficients in energy and angular momentum in the 
spherical case. 

The loss cone is taken into account by computing the radius of each periapse from the 
numerically integrated trajectory and eliminating the particle if this radius is 
below $\rcapt$. This automatically implies an appropriate boundary condition for the 
angular momentum diffusion in terms of $q$, which applies for both regular changes 
of $L$ due to precession in the smooth potential and the random two-body relaxation. 

In the present study, we neglect the changes in the background potential in the course 
of evolution. Most galactic nuclei have long enough relaxation times that the change in 
density profile due to relaxation in energy is negligible during the Hubble time 
(a few counterexamples exist, notably the Milky Way \citep{Merritt2009}). 
The change in shape which is responsible for the collisionless changes in angular 
momentum is more difficult to assess. Early studies \citep{GerhardBinney1985, 
MerrittQuinlan1998} have found that for a large enough black hole 
($\gtrsim 1\%$ of the galaxy mass) the triaxiality of the system is rapidly destroyed,
presumably, due to diffusion of chaotic orbits. However, later papers were successful 
at creating strongly triaxial models with high fraction of chaotic orbits which 
nevertheless maintained their shape during many dynamical times 
\citep{HolleyMSHN2002,PoonMerritt2004}. 
The changes in shape are quite straightforward to account for in our approach -- 
one needs to regularly update the smooth potential model as the system evolves; 
however, for the simplicity of interpretation we decided not to explore this 
possibility, having checked that the systems under study do not noticeably evolve 
in shape in the absense of relaxation and capture. 
(Noise can substantially increase the chaotic diffusion rate \citep[e.g.][]
{KandrupPS2000}, we leave the impact of noise on the chaotic orbits for a future study).

We prepared a number of test models, for three geometries (spherical, axisymmetric 
and triaxial), constructed by the \citet{Schwarzschild1979} orbit superposition method,
using the publically available \textsl{SMILE} software \citep{Vasiliev2013}.
A few modifications were introduced in order to enhance the quality of models. 
First, to increase the effective ``resolution'' of the orbit library, we have 
changed the scheme for generating initial conditions in such a way that it creates 
more orbits with lower energies (closer to the black hole) and more orbits with 
low angular momenta. The actual weights of orbits are assigned by the optimization 
procedure to satisfy the density and kinematic constraints while attaining the 
most uniform possible distribution of orbit weights; this scheme was modified to 
correctly account for non-equal priors for orbit weights. Second, to make the 
non-spherical models as close as possible to spherical in terms of distribution 
in angular momentum, we binned the distribution function in $\{E,\calR\}$ space 
on a non-uniform grid (with smaller cells closer to loss-cone boundary) and 
required the distribution in $\calR$ to be flat in each energy shell. 
This produces a more isotropic model in the angular momentum space than just 
a requirement of velocity isotropy. The models were constructed using $10^5$ 
orbits, $100\times10$ constraints for the spherical-harmonic decomposition of 
density and $100\times5$ constraints for angular momentum distribution. 
The increase in effective resolution allowed us to have $\sim$ factor of 10 more 
centrophilic orbits compared to a uniform sampling; accordingly, the typical 
weights of these orbits were also a factor of 10 less than average.

To validate the predictions of capture rates given by the Monte-Carlo code, 
we compared it to a number of direct $N$-body simulations, in the regime where 
the latter are feasible. As in our previous studies, we used the efficient 
direct $N$-body code $\phi$GRAPE-chain \citep{HarfstGMM2008}. 
It combines direct summation, using hardware acceleration of force computation 
provided by the \textsl{SAPPORO} library \citep{GaburovHP2009}, with chain 
regularization algorithm for accurate treatment of close encounters of stars 
with the black hole. The accuracy parameter of Hermite integrator was $\eta=0.01$ 
and the softening length was set to zero. 

As explained in the previous section, it is very difficult if not impossible 
to construct even a scaled $N$-body model for a realistic galactic nucleus.
Furthermore, for the difference between three geometries to be most prominent, 
we need to have a situation when the transition between empty and full loss cone 
regimes occurs well outside influence radius, and the draining time of saucer 
orbits at $\rh$ is much less than the simulation time (otherwise we don't expect 
to have a substantial difference between axisymmetric and triaxial cases -- both 
would have almost full loss cone draining rates at $\rh$). 
An extensive search through the parameter space has led us to adopt the following 
parameters for the comparison between $N$-body and Monte-Calso simulations:
$\gamma=1.5$ \citet{Dehnen1993} model with total mass of unity and $\Mbh=0.1$
(this gives $\rh\approx0.53$ and $\trad(\rh)\approx 1.7$), capture radius 
$\rcapt=2\times10^{-4}$, and simulation time of 20 $N$-body time units. 
The non-spherical models had axis ratios $1:1:0.75$ and $1:0.9:0.75$ for 
the axisymmetric and triaxial case, correspondingly, while having the same average
$M(r)$ profile as the spherical one. We stress again that these models do not 
correspond to any realistic galactic nuclei, but allow us to test the code.
We have set up three $N$-body simulations with $N=5\times10^5$ particles
using $\phi$GRAPE, and ran the Monte-Carlo code on the same initial snapshots, 
using the value of Coulomb logarithm $\ln\Lambda = 11$ (calibrated against the 
energy diffusion rate measured in the $N$-body simulation).

\begin{figure}
$$\includegraphics{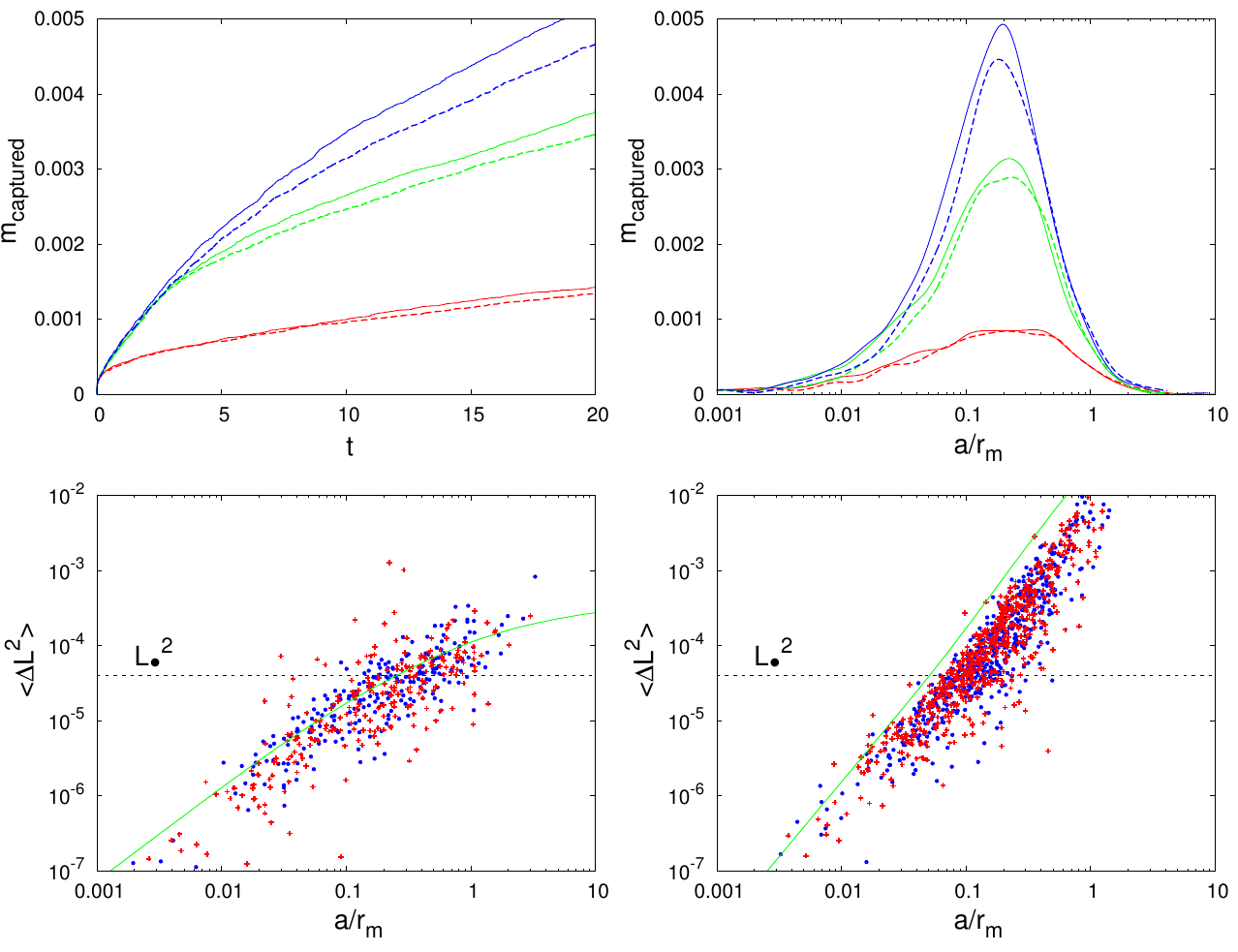}$$
\caption{Comparison between $N$-body (solid lines) and Monte-Carlo (dashed lines) 
simulations of a $N=5\times10^5$, $\gamma=3/2$ 
Dehnen model with $\Mbh=0.1$, for three cases: spherical (red, bottom lines), 
axisymmetric (green, middle lines) and triaxial (blue, top lines). 
Other parameters of the simulations are given in the text. \protect\\
\textit{Top left:} cumulative mass of captured particles as a function of time.
\textit{Top right:} distribution of captured particles in semimajor axis $a$, 
normalized by the influence radius ($\rh=0.53$ in model units).
\textit{Bottom row:} change in squared angular momentum $L^2$ in one radial period 
before capture, for the spherical (left) and triaxial (right) case; axisymmetric 
case is similar to triaxial. The transition between empty and full loss cone 
boundary conditions occurs when $\Delta L^2\approx \Lcapt^2$, and is shifted 
to smaller radii in non-spherical potentials. 
Red crosses are from $N$-body simulations, blue dots -- from Monte-Carlo models, 
green lines are the analytic trends \citep[see][Fig.8]{VasilievMerritt2013}.
} \label{fig:comparison_nb_mc_fp}
\end{figure}

Figure \ref{fig:comparison_nb_mc_fp} shows the comparison between $N$-body and 
Monte-Carlo codes for the three geometries. 
We considered the time dependence of the number of captured particles and 
their distribution in semimajor axis $a$.
The agreement between two codes is fairly good, with the $N$-body simulation 
giving slightly larger number of captured stars for $a \lesssim 0.2\rh$; 
this enhancement may be a result of resonant relaxation, which is not accounted 
for in the Monte-Carlo code, or large-angle deflections, which are also neglected. 
We also show the change in $L^2$ during one radial period just before capture;
as expected, this change is larger for non-spherical models, and the transition 
between empty and full loss cone regimes occurs well within $\rh$ for most 
realistic parameters \citep{VasilievMerritt2013}. 
We note that relativistic precession starts playing role for radii smaller than 
roughly the radius of transition to the full loss cone regime 
\citep{MerrittVasiliev2011}, and therefore our neglect of relativity does not 
lead to noticeable errors in the overall capture rate. 
As a final remark, the Monte-Carlo simulation of this scale takes only half 
an hour on a typical multi-core workstation, while the $N$-body simulation takes 
approximately 10 days with a modern GPU.

\section{Results}  \label{sec:results}

\begin{figure}
$$\includegraphics{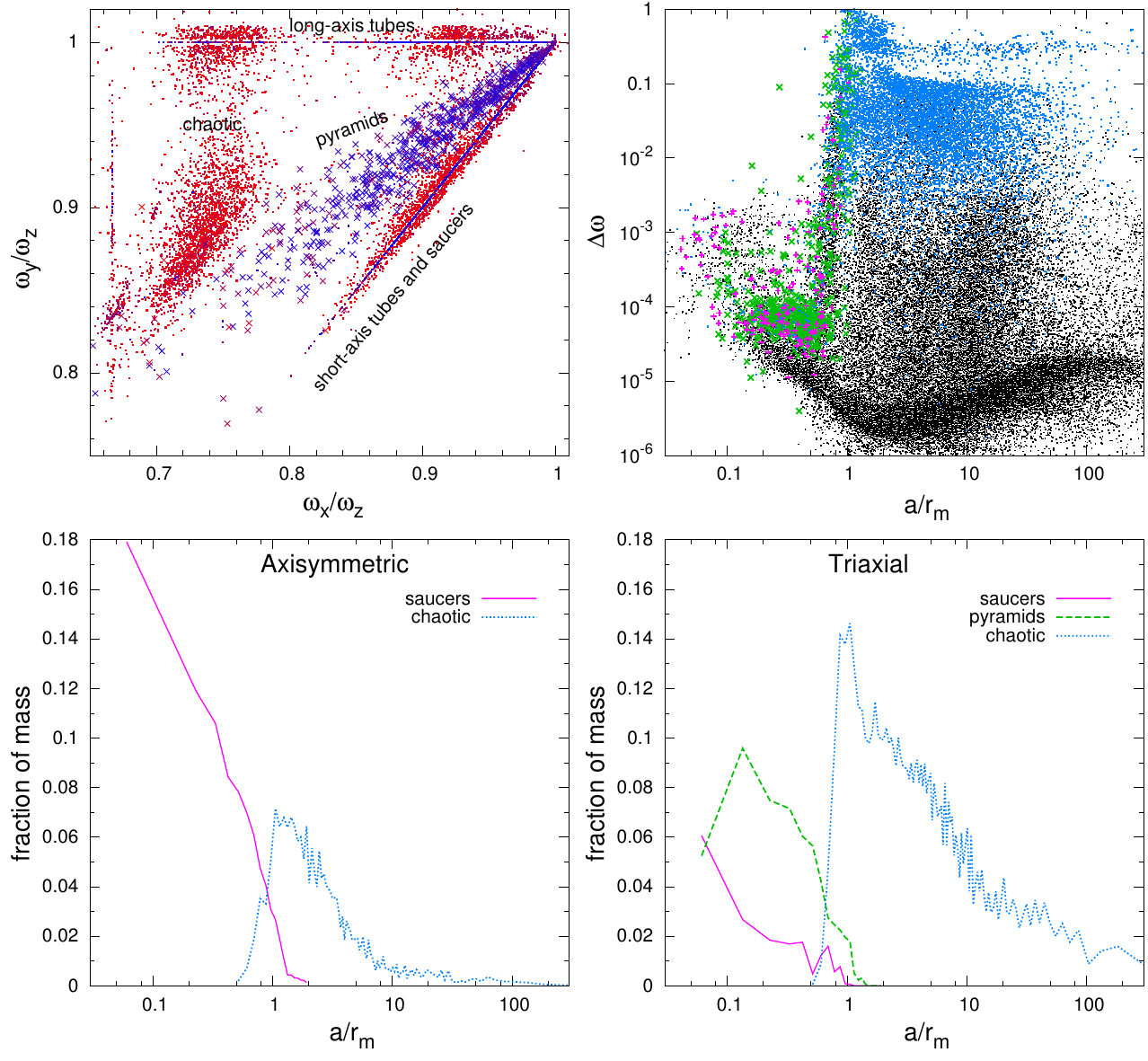}$$
\caption{Properties of axisymmetric and triaxial $\gamma=1$ Dehnen models 
with $\Mbh=10^{-2}$ and axis ratios $1:1:0.8$ and $1:0.9:0.8$. \protect\\
\textit{Top left:} frequency map of the triaxial model 
\citep[e.g.][]{ValluriMerritt1998}. 
Each point represents an orbit with the given ratio of most prominent frequencies 
of motion; Color indicates the degree of chaoticity, measured by the frequency 
diffusion rate (blue -- regular, red -- chaotic). Tube orbits are grouped along 
the lines $\omega_y=\omega_x$ (short-axis tubes and saucers) and $\omega_x=\omega_z$
(long-axis tubes); pyramid orbits (marked by crosses), which are regular, occupy the 
region between two tube families, sharing it with chaotic box orbits (red dots). 
\protect\\
\textit{Top right:} frequency diffusion rate as a function of semimajor axis $a$, 
normalized to the influence radius $\rh$, for the same triaxial model. 
This quantity measures the degree of chaos for each orbit 
\citep[see][\S4.2 for a discussion]{Vasiliev2013}.
Inside $\rh$ orbits are regular, some of them belong to families of pyramids 
(green diagonal crosses) or saucers (pink vertical crosses).
Between 0.5 and 1 $\rh$ there is a rapid transition to chaos in the orbit population; 
these chaotic orbits are also centrophilic (marked by blue dots) and usually have 
$\delta\omega \gtrsim 10^{-2}$, while the majority of orbits are regular, centrophobic 
(black dots) and have much lower $\Delta\omega$. 
The algorithm for detection of centrophilic orbits is described in the appendix of 
\citet{VasilievAM2014}. \protect\\
\textit{Bottom panels:} fraction of mass in various orbit families, as a function 
of semimajor axis normalized by $\rh$. 
Left -- axisymmetric model, right -- triaxial model.
} \label{fig:model_orbits}
\end{figure}

\begin{figure}
$$\includegraphics{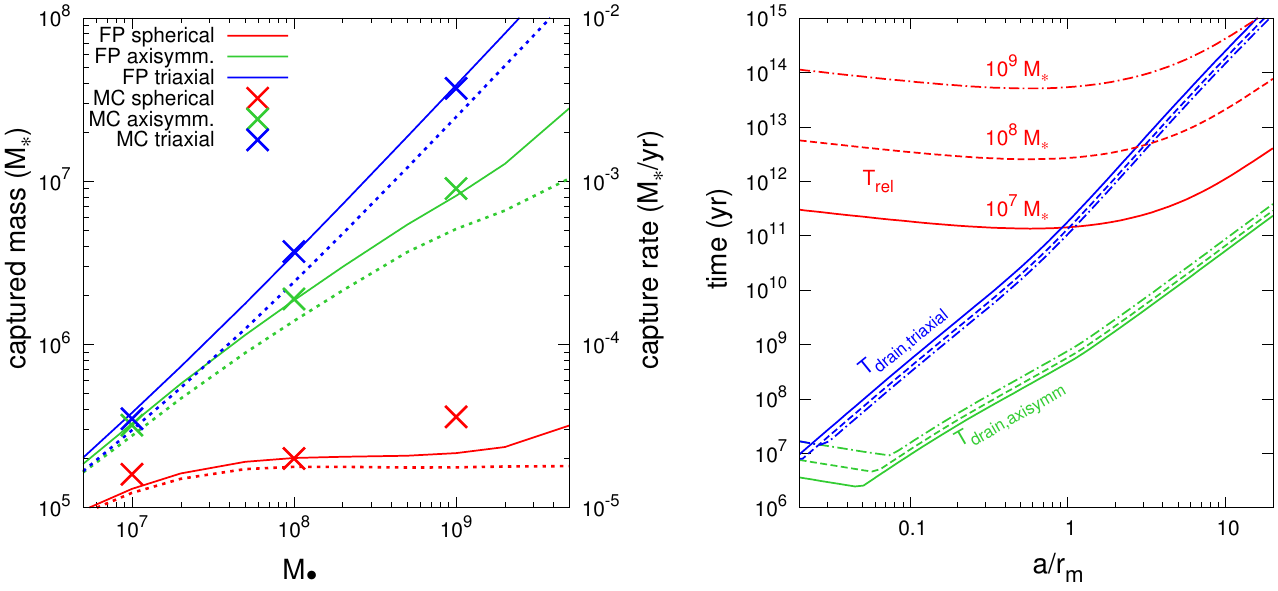}$$
\caption{
\textit{Left panel:} The total mass of captured stars after $10^{10}$ years
(left axis), and the present-day capture rate (right axis), as functions of 
black hole mass $\Mbh$. 
The crosses show the total captured mass in Monte-Carlo simulations for three 
geometries; solid lines show the same quantity computed using time-dependent 
Fokker-Planck models from \citet{VasilievMerritt2013}, which 
include the contribution of draining; and dotted lines show the present-day 
capture rates derived from the same models. 
For the triaxial case, the capture rate is almost entirely due to draining 
of centrophilic orbits (Equation~\ref{eq:Ftotal_triax}).
\protect\\
\textit{Right panel:} Various timescales as functions of radius (normalized to 
the influence radius $\rh$). 
Shown is the local relaxation time (red, top), draining time of centrophilic 
orbits in the triaxial case (blue, middle, Eq.~\ref{eq:tdrain_tri}) 
and in the axisymmetric case (green, bottom, Eq.~\ref{eq:tdrain_axi}).
Solid lines -- $\Mbh=10^7\,M_\odot,\ \rh=13$~pc, dashed -- $\Mbh=10^8\,M_\odot,
\ \rh=45$~pc, dot-dashed -- $\Mbh=10^9\,M_\odot,\ \rh=160$~pc.
} \label{fig:summary}
\end{figure}

\begin{figure}
$$\includegraphics{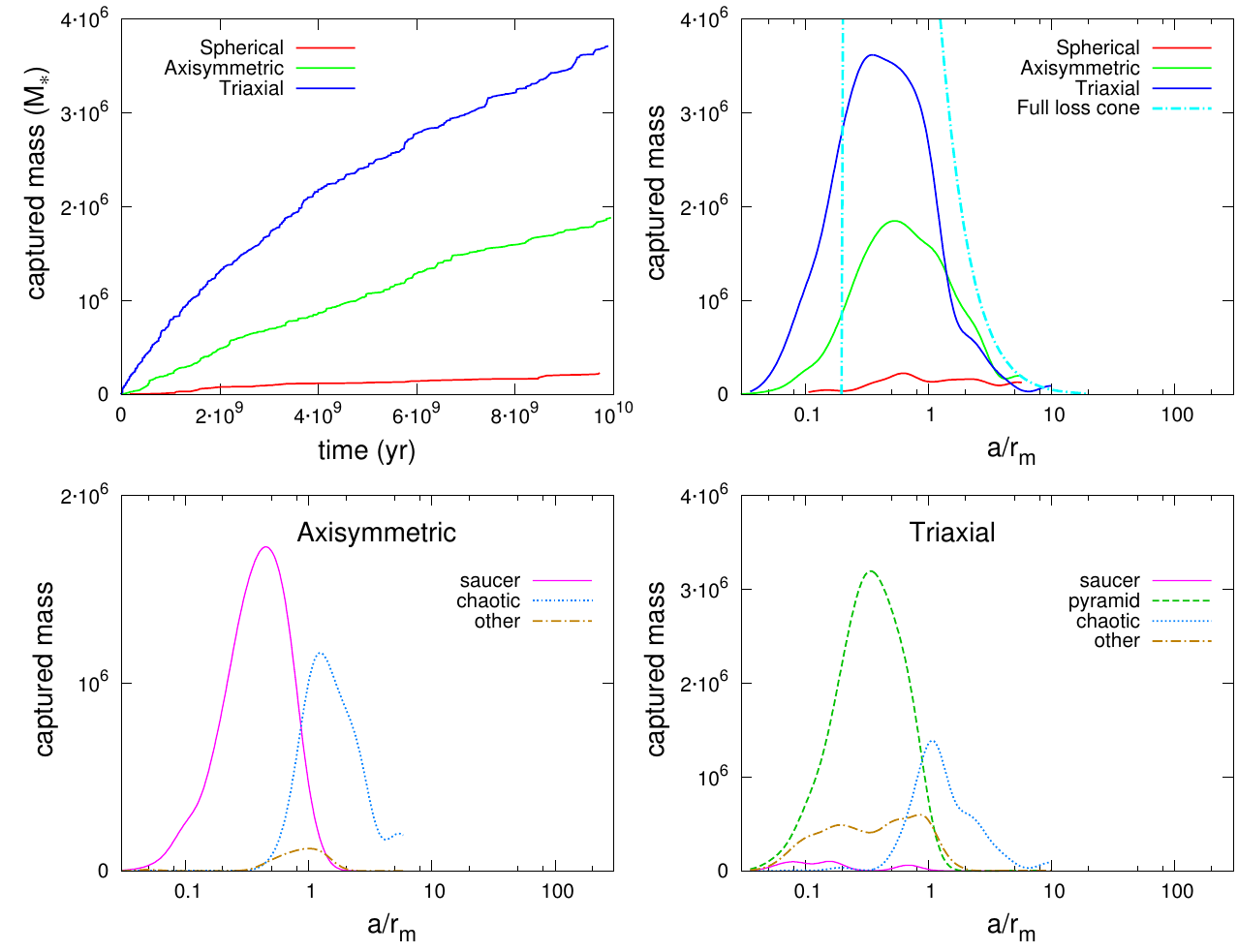}$$
\caption{Properties of captured stars for a $\Mbh=10^8$ black hole.\protect\\
\textit{Top left:} cumulative mass as a function of time, for three models 
(from top to bottom: triaxial, axisymmetric, spherical). \protect\\
\textit{Top right:} distribution of captured stars in semimajor axis $a$, 
normalized by the influence radius $\rh$. In addition to the same three models, 
the full loss cone rate (\ref{eq:flux_full_lc}) is plotted as dot-dashed line, 
and the ``draining radius'', at which the draining time (\ref{eq:tdrain_tri}) 
equals $10^{10}$~yr, is marked by a vertical line.\protect\\
\textit{Bottom panels:} distribution of captured stars in $a/\rh$ for each orbit type.
} \label{fig:captured}
\end{figure}

Having tested the Monte-Carlo code on a ``toy'' model, we now apply it for the models 
describing realistic galactic nuclei. We take the Dehnen density profile with 
$\gamma=1$, a black hole mass of either $10^{-2}$ or $10^{-3}$ of the galaxy mass, 
and axis ratios of $1:1:0.8$ and $1:0.9:0.8$, corresponding to $\eaxi\approx 0.27$. 
The orbit composition of models is shown on Figure~\ref{fig:model_orbits}.
The fraction of ``orbits of interest'' -- saucers, pyramids and chaotic orbits -- 
is $\sim 10-20$\% of mass near $\rh$, but their total mass is lower: 
the mass of chaotic orbits is $\sim 1.5\,(5)\%$ of the total galaxy mass and 
the mass of saucers (pyramids) is $\sim 16\,(5)\%\Mbh$ for the axisymmetric 
(triaxial) models. Thanks to the improved mass sampling scheme, the number fraction
of these orbits is much larger than their mass fraction.
The models are scaled to physical scale using three values of 
$\Mbh = 10^7, 10^8$ and $10^9\,M_\odot$, 
capture radius $\rcapt = 4\times 10^{-5} (\Mbh/10^8\,M_\odot)$~pc 
and the influence radius $\rh = 45\, (\Mbh/10^8\,M_\odot)^{0.56}$~pc. 
The simulation time is $10^{10}$~yr and the dynamical time at the radius of influence 
is $\sim 10^6\,(\Mbh/10^8\,M_\odot)^{0.34}$~yr. 
We do not include physical collisions in our treatment, since for our models 
the collision timescale is much longer than the Hubble time at all radii of interest, 
although the situation could well be different for less massive black holes in dense 
nuclei \citep{DuncanShapiro1983,MurphyCD1991,FreitagBenz2002}. Likewise, 
the relaxation time is also long enough to neglect the changes in density profile.
We consider a population of equal-mass main-sequence stars with solar mass and radius; 
this is surely a crude simplification, as the red giants have much larger physical 
radii and are more readily torn apart by tidal forces \citep{SyerUlmer1999}. 
Nevertheless, the basic model should capture qualitatively the main features of more 
realistic galactic nuclei.

The results of Monte-Carlo simulations agree well with the qualitative picture 
described above. The difference between three geometries is modest at low $\Mbh$, 
when even the spherical system is largely in the full loss cone regime, but 
becomes more pronounced for $\Mbh \gtrsim 10^8\,M_\odot$ 
(Figure~\ref{fig:summary}).
The peak of distribution of captured stars in energy corresponds to semimajor axis 
$a\sim \rh/2$ (as the orbits of captured stars are almost radial, one may say that 
these stars come from radii $\sim 2a \sim \rh$). 
At this energy, the draining time of loss region is $\sim4\times 10^{10}$~yr for 
the triaxial model and $\sim 3\times 10^8$~yr for the axisymmetric one, in both cases 
weakly depending on $\Mbh$. 
Thus in spherical and axisymmetric systems the flux at present time is mostly 
determined by relaxation, while for triaxial systems it is dominated by draining of 
centrophilic orbits (mostly pyramids within $\rh$). The difference between the first 
two comes from the larger size of the effective capture boundary in the axisymmetric 
case, which enhances the flux by a logarithmic factor of few ($\sim 
\ln(\calRcapt^{-1})/\ln(\eaxi^{-1})$, \citet{VasilievMerritt2013}), but only in the 
case when most of the flux in the spherical case occurs in the empty loss cone regime. 
Again, one should keep in mind that this refers to the steady-state flux, while 
in the above figure we plot the solution of the time-dependent Fokker-Planck equation, 
in which the difference between spherical and axisymmetric systems is much larger 
if the evolution time is shorter than $\sim 0.1\trel(\rh)$ (i.e., for 
$\Mbh\gtrsim 10^7\,M_\odot$).

Figure~\ref{fig:captured} shows the properties of captured stars for a $10^8\,M_\odot$ 
black hole in three geometries. In this case the difference in the total captured mass 
between spherical and axisymmetric cases is already an order of magnitude, while 
the triaxial system adds another factor of two. The figure also shows that, contrary 
to a naive expectation, the flux at almost all energies is substantially lower than 
the full loss cone flux; nevertheless, the crude estimate based on a full loss cone 
rate above the ``draining'' energy (\ref{eq:Ftotal_triax}) is remarkably close to the 
more refined calculation. Chaotic orbits contribute a minor fraction to the captured 
mass in both axisymmetric and triaxial cases.
The figure was plotted for the model with $\Mbh=10^{-2}$ of the total galaxy mass; 
the model with $\Mbh=10^{-3}$ has very similar properties (to within 10\%) but 
less reliable statistics, as the number of captured orbits is lower by a factor of 3.

An important question about non-spherical models with black holes is whether the 
scattering of chaotic orbits by the central mass may lead to the elimination or 
reduction of triaxiality. We do not see any evidence for this, even though the models 
were evolved for $\sim 10^4$ dynamical times at the radius of influence. 
The opposite conclusion reached in earlier studies 
\citep{GerhardBinney1985,MerrittQuinlan1998,Kalapotharakos2008} might have resulted from 
considering evolutionary rather than steady-state models, in which a black hole 
grew in a pre-existing triaxial cusp.

We have also considered a model with a larger deviation from spherical symmetry, 
namely with axis ratio $1:0.8:0.5$, same as in \citet{PoonMerritt2004}, and the 
black hole mass of 1\% of galaxy mass. 
However, it turned out to be impossible to create a quasi-isotropic model with 
these properties: there is no difficulty to arrange for the given shape and 
density profile, but the region outside a few $\rh$ cannot have a uniform 
distribution of stars in $\calR$. Instead a radially anisotropic system is 
formed with a scarcity of orbits at intermediate $\calR$ and an excess of them 
at $\calR=1$ and even more at $\calR=0$. Naturally, such system may have a draining 
rate larger than the full loss cone rate, which might explain the conclusions reached 
by \citet{MerrittPoon2004}. Instead, we opted for a model with variable triaxiality: 
the axis ratio $y/x$ increased from 0.8 to 0.9 between 5 and 7$\,\rh$, i.e.\ outside 
the region that gives the main contribution to the total flux. 
As expected, this model yielded a larger captured mass, roughly by a factor 1.5.

We also confirmed the scaling relations derived in \S\ref{sec:scaling}, by performing 
a simulation with $\rcapt, \calD$ and $t_\mathrm{sim}^{-1}$ all scaled by the same 
factor of 10. This resulted in a very similar behaviour for axisymmetric and triaxial 
systems, while somewhat increasing the flux in the spherical case 
(due to a logarithmic dependence on $\calRcapt$).

\section{Conclusions}  \label{sec:conclusions}

We have considered the dynamics of stars in galactic nuclei containing supermassive 
black holes, with the focus on the rates of capture of stars by the black hole, 
in particular, in non-spherical nuclei. 
The supply of stars into the loss cone -- the region of phase space from where the 
capture is possible -- is driven by changes in angular momentum due to two effects:
collisional two-body relaxation, which operates in any geometry, and collisionless 
torques which arise in the case that the potential is non-spherical.

We have reviewed the structure of orbits in galactic nuclei and identified the classes 
of orbits that are important for the loss cone problem -- saucer orbits in axisymmetric 
systems and pyramid orbits in triaxial systems, as well as chaotic orbits which exist 
outside the black hole radius of influence $\rh$ in both cases. These orbits form the
loss region, from which a star may get into the loss cone and be captured by the black 
hole even in the absence of two-body relaxation, due to collisionless torques only.
In a triaxial case, chaotic and pyramid orbits are centrophilic, i.e.\ they may attain 
arbitrarily low values of angular momentum; in an axisymmetric case, the loss region 
is composed of orbits with the conserved component of angular momentum, $L_z$, being 
low enough to let them enter the loss cone. In both cases the volume of the loss region 
is substantially larger than the loss cone proper, thus we expect a higher capture rate 
due to two factors: (i) the stars in the loss region are drained over a timescale 
much longer than the dynamical time, and (ii) two-body relaxation is required only to 
drive a star into the loss region, instead of a much smaller loss cone, for it to be 
eventually captured. 

The interplay between collisonless and collisional effects is poorly handled by 
present-day $N$-body simulations, due to the necessity of a very large number of 
particles for the collisional effects to be realistically small. To circumvent this 
trouble, we have developed a Monte-Carlo method that can simulate these systems with 
a rather modest number of particles while ensuring a faithful balance between the two 
effects. It follows the orbits of stars in a smooth potential of any geometry while 
perturbing them according to the standard two-body relaxation theory. 
We applied this method to the loss-cone problem in non-spherical galaxies, with the aim 
of determining the total mass of stars captured by the black hole during the Hubble 
time. We assumed a one-parameter family of models with a $\rho\propto r^{-1}$ density 
cusp and the scale radius of the model determied by the $\Mbh-\sigma$ relation. 
Our main results are the following:
\begin{itemize}
\item Most stars find their way into the black hole from regular orbits inside $\rh$. 
In a triaxial system, the timescale for collisionless draining of centrophilic orbits 
is larger than the galaxy age, unless the triaxiality is very small. Thus the bulk 
of the capture events are not caused by two-body encounters, but are due to the 
torques from the smooth non-spherical mass distribution.
In an axisymmetric system, the draining time is short compared to the galaxy age, 
but the larger size of the loss region compared to the loss cone proper means that 
the overall capture rate due to collisional relaxation is increased by a factor of few 
with respect to the spherical case. 
The difference between three geometries is larger for larger black holes, being more 
than a factor of ten for $\Mbh \gtrsim 10^8\,M_\odot$, but only a factor of two 
for $\Mbh \lesssim 10^7\,M_\odot$. 
Thus we have confirmed earlier studies \citep[e.g.][]{MagorrianTremaine1999,
MerrittPoon2004} that have suggested the importance of draining of the loss region 
for the capture rates in non-spherical galaxies.
\item The initial draining rate for a system with a uniform distribution in $\calR$ 
(i.e.\ isotropic in velocity space) is roughly equal to the full loss cone rate. 
However, as the draining time is strongly dependent on energy, the total draining flux, 
integrated over all energies, decreases with time, as the most bound orbits are 
progressively depleted. 
There is no globally well-defined full loss cone rate, but even the value of full loss 
cone flux at the radius of influence overestimates the actual present-day flux 
by a factor of few. A similar conclusion was reached by \citet{VasilievAM2014} for 
the case of a binary black hole in a triaxial galaxy.
The draining of centrophilic orbits in the triaxial system may account for the total 
captured mass of a few percent of $\Mbh$, scaling roughly linearly with the latter 
and weakly depending on the degree of triaxiality. 
\item In non-spherical systems, stars in the loss cone have a more uniform distribution 
in periapsis radius than in the spherical case; in other words, the boundary condition 
corresponds to the full loss cone regime for most captured orbits. This could have 
observational implications for the properties of the tidal disruption flares 
\citep{StrubbeQuataert2009}, although recent studies suggest that the value of impact 
parameter has little influence on the dynamics of the tidal disruption event 
\citep{StoneSL2013, GuillochonRuiz2013}.
\item Chaos in the population of low angular momentum orbits, induced by the central 
black hole, does not necessarily lead to the loss of triaxiality over the lifetime 
of the galaxy. Our models, set up initially in nearly perfect equilibrium, retained 
their shape for $\sim 10^4$ dynamical times at $\rh$ (for a $10^8\,M_\odot$ black hole).
\end{itemize}

These conclusions were confirmed for a limited set of models but are likely to hold 
for more general cases. The most important assumption in the present study was that 
the initial distribution of stars was close to isotropic in the angular momentum space, 
which results in a rather large draining flux, and an increased capture rate due to 
relaxation, compared to the steady-state value. If a galaxy has an already depleted 
stellar population at low angular momenta, for instance, due to ejection of these 
orbits by a binary supermassive black hole \citep{MerrittWang2005}, then the difference 
between the three geometries is likely to be reduced, and the overall flux will be lower.
Thus our calculations should be taken rather as upper estimates, which nevertheless 
predict a rather high capture rate ($\sim10^{-3}\,M_\odot/$yr) for the most massive black
holes in non-spherical galaxies. On the other hand, the number of detectable tidal 
disruption flares for these black holes is much smaller, as for black holes more massive 
than $10^8\,M_\odot$ most of the captured stars are swallowed entirely without producing 
any flare. Figure~15 in \citet{MacLeodGR2012} suggests that for $\Mbh=10^9\,M_\odot$ 
the tidal disruption flares of giant stars could constitute less than 5\% of total 
capture events. A more quantitative estimate is beyond the scope of this paper, 
we only note for more realistic assumptions about stellar structure and evolution there 
are several other processes that could be important in the context of star--black hole 
interaction, such as ``growth'' of a giant star into the loss cone \citep{SyerUlmer1999,
MacLeodRGG2013}, or tidal dissipation--induced capture \citep{LiLoeb2013}.
It is also worth noting that we considered rather low-density galactic nuclei, in which 
we could neglect physical collisions between stars and the changes in density profile 
due to relaxation in energy, both processes occuring on timescales much longer than 
Hubble time. For less massive black holes in denser nuclei (like the Milky Way) the 
situation is more complicated, but in these systems we don't expect non-spherical effects 
to play a significant role in the capture rate.

I am grateful to the referees for their valuable comments which helped to improve 
the presentation.
This work was partly supported by the National Aeronautics and Space Administration 
under grant no. NNX13AG92G. 
The software for estimating capture rates from Fokker-Planck models, including draining 
of the loss region, is available from 
\url{http://td.lpi.ru/~eugvas/losscone/}.

\def\newblock{}
\setlength{\bibsep}{0pt plus 0.3ex}


\begin{thebibliography}{}
\footnotesize
\bibitem[Amaro-Seoane et al.(2004)]{AmaroSeoaneFS2004}
Amaro-Seoane P., Freitag M., \& Spurzem R.\ 2004, MNRAS, 352, 655

\bibitem[Antonini(2014)]{Antonini2014}
Antonini, F.\ 2014, ApJ, 794, 106

\bibitem[Bahcall \& Wolf(1976)]{BahcallWolf1976} 
Bahcall, J. \& Wolf, R.\ 1976, ApJ, 209, 214

\bibitem[Baumgardt et al.(2004)]{BaumgardtME2004}
Baumgardt H., Makino J., Ebisuzaki T.\ 2004, ApJ, 613, 1133

\bibitem[Binney \& Tremaine(2008)]{BT2008}
Binney, J., \& Tremaine, S.\ 2008, Galactic dynamics (Princeton: Princeton University Press)

\bibitem[Brockamp et al.(2011)]{BrockampBK2011}
Brockamp, M., Baumgardt, H., \& Kroupa, P.\ 2011, MNRAS, 418, 1308

\bibitem[Cohn \& Kulsrud(1978)]{CohnKulsrud1978} 
Cohn, H., \& Kulsrud, R.\ 1978, ApJ, 226, 1087

\bibitem[Dehnen(1993)]{Dehnen1993} 
Dehnen, W.\ 1993, MNRAS, 265, 250 

\bibitem[Duncan \& Shapiro(1983)]{DuncanShapiro1983}
Duncan, M., \& Shapiro, S.\ 1983, ApJ, 268, 565

\bibitem[Ferrarese \& Merritt(2000)]{FerrareseMerritt2000} 
Ferrarese, L., \& Merritt, D.\ 2000, ApJ, 539, L9 

\bibitem[Fiestas \& Spurzem(2010)]{FiestasSpurzem2010}
Fiestas, J., \& Spurzem, R.\ 2010, MNRAS, 405, 194

\bibitem[Fiestas et al.(2012)]{FiestasPBS2012}
Fiestas, J., Porth, O., Berczik, P., \& Spurzem, R.\ 2012, MNRAS, 419, 57

\bibitem[Frank \& Rees(1976)]{FrankRees1976}
Frank, J., \& Rees, M.\ 1976, MNRAS, 176, 633

\bibitem[Freitag \& Benz(2002)]{FreitagBenz2002}
Freitag, M., \& Benz, W.\ 2002, A\&A, 394, 345

\bibitem[Gaburov et al.(2009)]{GaburovHP2009}
Gaburov, E., Harfst, S., Portegies Zwart, S.\ 2009, New Astron., 14, 630

\bibitem[Gebhardt et al.(2000)]{Gebhardt2000} 
Gebhardt, K., Bender, R., Bower, G., et al.\ 2000, ApJ, 539, L13 

\bibitem[Gerhard \& Binney(1985)]{GerhardBinney1985}
Gerhard, O., \& Binney, J.\ 1985, MNRAS, 216, 467

\bibitem[Giersz(1998)]{Giersz1998}
Giersz, M.\ 1998, MNRAS, 298, 1239

\bibitem[Goodman(1983)]{Goodman1983}
Goodman, J.\ 1983, PhD Thesis, Princeton Univ.

\bibitem[Guillochon \& Ramirez-Ruiz(2013)]{GuillochonRuiz2013}
Guillochon, J., \& Ramirez-Ruiz, E.\ 2013, ApJ, 767, 25

\bibitem[G\"ultekin et al.(2009)]{Gultekin2009}
G\"ultekin, K., Richstone, D., Gebhardt, K., et al.\ 2009, ApJ, 698, 198

\bibitem[Harfst et al.(2008)]{HarfstGMM2008} 
Harfst, S., Gualandris, A., Merritt, D., \& Mikkola, S.\ 2008, MNRAS, 389, 2

\bibitem[Henon(1971)]{Henon1971}
H\'enon, M. H., 1971, Ap\&SS, 13, 284

\bibitem[Holley-Bockelmann et al.(2002)]{HolleyMSHN2002}
Holley-Bockelmann K., Mihos J. C., Sigurdsson S., Hernquist L., Norman C., 2002, ApJ, 567, 817

\bibitem[Holley-Bockelmann \& Sigurdsson(2006)]{HolleySigurdsson2006}
Holley-Bockelmann, K., \& Sigurdsson, S.\ 2006, arXiv:astro-ph/0601520

\bibitem[Hopman \& Alexander(2006)]{HopmanAlexander2006} 
Hopman, C., \& Alexander, T. \ 2006, ApJ, 645, 1152

\bibitem[Joshi et al.(2000)]{JoshiRPZ2000}
Joshi, K. J., Rasio, F. A., \& Portegies Zwart, S. 2000, ApJ, 540, 969

\bibitem[Kalapotharakos(2008)]{Kalapotharakos2008}
Kalapotharakos, C., \ 2008, MNRAS, 389, 1709

\bibitem[Kandrup et al.(2000)]{KandrupPS2000}
Kandrup, H., Pogorelov, I., \& Sideris, I.\ 2000, MNRAS, 311, 719

\bibitem[Li \& Loeb(2013)]{LiLoeb2013}
Li, G., \& Loeb, A.\ 2013, MNRAS, 429, 3040

\bibitem[Lightman \& Shapiro(1977)]{LightmanShapiro1977}
Lightman, A., \& Shapiro, S.\ 1977, ApJ, 211, 244

\bibitem[MacLeod et al.(2012)]{MacLeodGR2012}
MacLeod, M., Guillochon, J., \& Ramirez-Ruiz, E.\ 2012, ApJ, 757, 134

\bibitem[MacLeod et al.(2013)]{MacLeodRGG2013}
MacLeod, M., Ramirez-Ruiz, E., Grady, S., \& Guillochon, J. \ 2013, ApJ, 777, 133

\bibitem[Magorrian \& Tremaine(1999)]{MagorrianTremaine1999}
Magorrian J., \& Tremaine S.\ 1999, MNRAS, 309, 447

\bibitem[M\'endez-Abreu et al.(2010)]{MendezAbreu2010}
M\'endez-Abreu, J., Simonneau, E., Aguerri, J., \& Corsini, E. \ 2010, A\&A, 521, A71

\bibitem[Merritt(2009)]{Merritt2009} 
Merritt, D.,\ 2009, ApJ, 694, 959

\bibitem[Merritt(2013a)]{Merritt2013} 
Merritt, D.,\ 2013a, Clas.Quant.Grav., 30, 244005

\bibitem[Merritt(2013b)]{MerrittBook} 
Merritt, D., 2013b, Dynamics and evolution of galactic nuclei (Princeton: Princeton University Press)

\bibitem[Merritt et al.(2011)]{MerrittAMW2011}
Merritt, D., Alexander, T., Mikkola, S., Will, C.\ 2011, Phys.Rev.D, 84, 044024

\bibitem[Merritt \& Poon(2004)]{MerrittPoon2004} 
Merritt, D., \& Poon, M.\ 2004, ApJ, 606, 788 

\bibitem[Merritt \& Quinlan(1998)]{MerrittQuinlan1998} 
Merritt, D., \& Quinlan, G.~D.\ 1998, ApJ, 498, 625 

\bibitem[Merritt \& Valluri(1999)]{MerrittValluri1999} 
Merritt, D., \& Valluri, M.\ 1999, AJ, 118, 1177

\bibitem[Merritt \& Vasiliev(2011)]{MerrittVasiliev2011} 
Merritt, D., \& Vasiliev, E.\ 2011, ApJ, 726, 61

\bibitem[Merritt \& Wang(2005)]{MerrittWang2005} 
Merritt, D., \& Wang, J.\ 2005, ApJL, 621, L101 

\bibitem[Milosavljevi{\'c} \& Merritt(2003)]{MilosMerritt2003} 
Milosavljevi{\'c}, M., \& Merritt, D.  2003, ApJ, 596, 860

\bibitem[Murphy et al.(1991)]{MurphyCD1991}
Murphy, B., Cohn, H., \& Durisen, R.  1991, ApJ, 370, 60

\bibitem[Norman \& Silk(1983)]{NormanSilk1983}
Norman, C., \& Silk, J.\ 1983, ApJ, 266, 502

\bibitem[Perets et al.(2007)]{PeretsHA2007}
Perets, H., Hopman, C., Alexander, T.\ 2007, ApJ, 656, 709

\bibitem[Poon \& Merritt(2004)]{PoonMerritt2004} 
Poon, M.~Y., \& Merritt, D.\ 2004, ApJ, 606, 774 

\bibitem[Rauch \& Tremaine(1996)]{RauchTremaine1996} 
Rauch, K., \& Tremaine, S.\ 1996, New Astron., 1, 149

\bibitem[Rees(1988)]{Rees1988}
Rees, M.\ 1988, Nature, 333, 523

\bibitem[Sambhus \& Sridhar(2000)]{SambhusSridhar2000} 
Sambhus, N., \& Sridhar, S.\ 2000, ApJ, 542, 143

\bibitem[Sch\"odel et al.(2014)]{Schoedel2014}
Sch\"odel, R., Feldmeier, A., Kunneriath, D., et al.\ 2014, A\&A, 566, 47

\bibitem[Schwarzschild(1979)]{Schwarzschild1979} 
Schwarzschild, M.\ 1979, ApJ, 232, 236

\bibitem[Shapiro \& Marchant(1978)]{ShapiroMarchant1978}
Shapiro, S. L., \& Marchant, A. B. 1978, ApJ, 225, 603

\bibitem[Spitzer \& Hart(1971)]{SpitzerHart1971}
Spitzer, L., \& Hart, M.\ 1971, ApJ, 164, 399

\bibitem[Spitzer \& Shapiro(1972)]{SpitzerShapiro1972}
Spitzer, L., \& Shapiro, S. L.\ 1972, ApJ, 173, 529

\bibitem[Sridhar \& Touma(1997)]{SridharTouma1997} 
Sridhar, S., \& Touma, J.\ 1997, MNRAS, 292, 657

\bibitem[Sridhar \& Touma(1999)]{SridharTouma1999} 
Sridhar, S., \& Touma, J.\ 1999, MNRAS, 303, 483

\bibitem[Stone et al.(2013)]{StoneSL2013}
Stone, N., Sari, R., \& Loeb, A.,\ 2013, MNRAS, 435, 809

\bibitem[Strubbe \& Quataert(2009)]{StrubbeQuataert2009}
Strubbe, L. \& Quataert, E.\ 2009, MNRAS, 400, 2070

\bibitem[Syer \& Ulmer(1999)]{SyerUlmer1999}
Syer, D., \& Ulmer, A.\ 1999, MNRAS, 306, 35

\bibitem[Tremblay \& Merritt(1996)]{TremblayMerritt1996}
Tremblay, B., \& Merritt, D.\ 1996, AJ, 111, 2243

\bibitem[Valluri \& Merritt(1998)]{ValluriMerritt1998} 
Valluri, M., \& Merritt, D.\ 1998, ApJ, 506, 686

\bibitem[van Velzen et al.(2011)]{Velzen2011}
van Velzen, S., Farrar, G., Gezari, S., et al.\ 2011, ApJ, 741, 73

\bibitem[Vasiliev(2013)]{Vasiliev2013}
Vasiliev, E.\ 2013, MNRAS, 434, 3174

\bibitem[Vasiliev(2015)]{Vasiliev2014}
Vasiliev, E.\ 2015, MNRAS, 446, 3150

\bibitem[Vasiliev \& Merritt(2013)]{VasilievMerritt2013}
Vasiliev, E., \& Merritt, D.\ 2013, ApJ, 774, 87

\bibitem[Vasiliev et al.(2014)]{VasilievAM2014}
Vasiliev, E., Antonini, F., \& Merritt, D.\ 2014, ApJ, 785, 163

\bibitem[Wang \& Merritt(2004)]{WangMerritt2004}
Wang, J., \& Merritt, D.\ 2004, ApJ, 600, 149

\bibitem[Zhao et al.(2002)]{ZhaoHR2002}
Zhao, H.-S., Haehnelt, M., \& Rees, M.\ 2002, New Astron., 7, 385

\end{thebibliography}
\end{document}